\documentclass[aps,prl,twocolumn,superscriptaddress,showpacs,10pt,amsmath,amssymb]{revtex4-1}
\usepackage{graphicx}
\usepackage{xcolor}

\date{\today}

\begin{document}

\title{Confirmation of the random tiling hypothesis for a decagonal quasicrystal}

\author{Alexander Kiselev}
\affiliation{Institut f\"ur Theoretische und Angewandte Physik, Universit\"at Stuttgart, Pfaffenwaldring 57, 70550 Stuttgart, Germany}

\author{Michael Engel}
\affiliation{Department of Chemical Engineering, University of Michigan, Ann Arbor, MI 48109, USA}

\author{Hans-Rainer Trebin}
\affiliation{Institut f\"ur Theoretische und Angewandte Physik, Universit\"at Stuttgart, Pfaffenwaldring 57, 70550 Stuttgart, Germany}

\begin{abstract}
Mechanisms that stabilize quasicrystals are much discussed but not finally resolved. We confirm the random tiling hypothesis and its predictions in a fully atomistic decagonal quasicrystal model by calculating the free energy and the phason elastic constants over a wide range of temperatures. The Frenkel-Ladd method is applied for the phonon part and an approach of uncorrelated phason flips for the configurational part. When lowering the temperature, a phase transition to an approximant occurs. Close to the transition temperature one of the phason elastic constants becomes soft.
\end{abstract}

\maketitle

Quasicrystals are among the most complex ordered solid phases found in condensed matter. Their constituent atoms possess many different local environments and are arranged in a tiling~\cite{levine_quasicrystals:_1984}. While perfect quasiperiodic tilings like the Penrose tiling obey strict matching rules of the tiles, thermal matching rule violations can disorder the tiling~\cite{elser_comment_1985,henley_random_1988}. Understanding the role of matching rule violations for the stabilization of quasicrystals constitutes an important fundamental problem, which so far has received attention only in idealized model systems~\cite{widom_transfer-matrix_1989, strandburg_phason_1989, tang_equilibrium_1990, tang_random-tiling_1990, shaw_long-range_1991, jeong_finite-temperature_1993}.

The ensemble of matching rule violations is described by an order parameter, the phason strain $\chi$, which corresponds to the deviation from the perfect quasiperiodic tiling. For small values of the phason strain, the number of matching rule violations grows linearly, and it has been suggested that the free energy is non-analytic, $F\propto |\chi|$~\cite{socolar_phonons_1986, DiVincenzo-Steinhardt_inbook_1990}. The result is a `locked' quasicrystalline ground state, which is \emph{stabilized energetically}, \textit{i.e.}~by minimizing the number of energetically costly matching rule violations. A locked state has been confirmed with Monte Carlo (MC) simulations in tiling models with matching rule based interactions at $T=0$ in two dimensions~\cite{tang_equilibrium_1990} and at $T>0$ in three dimensions~\cite{jeong_finite-temperature_1993}.

In contrast, if fluctuations of phason strain are present in equilibrium (`unlocked'), then a quasicrystal can be \emph{stabilized entropically}. This scenario is known as the random tiling hypothesis of Elser~\cite{elser_comment_1985} and Henley~\cite{henley_random_1988, Henley_randomtilingmodels_1990}. The free energy then follows a generalized elastic theory and is a quadratic form in phason strain with a minimum at the on average unstrained tiling, $F\propto\chi^2$. With decreasing temperature, one of the phason elastic constants can become soft, and a phase transition to a periodic approximant occurs~\cite{Henley_randomtilingmodels_1990}. The random tiling hypothesis has been investigated amply in simulations with monodisperse~\cite{tang_random-tiling_1990,shaw_long-range_1991} and binary tilings~\cite{strandburg_phason_1989} and with transfer matrix methods~\cite{henley_random_1988,widom_transfer-matrix_1989}. Random tiling quasicrystals have been identified in experiment, but it is an open question whether they can be ground states of matter~\cite{Steurer_phasetransformations_2005}. Besides quasicrystals, random tilings have been reported in packings of hard dimers~\cite{wojciechowski_nonperiodic_1991} and, recently, in molecular networks absorbed on graphite \cite{blunt_random_2008}.

Phonon contributions to the free energy of tilings were first taken into account by Koschella \textit{et al.} for the binary decagonal Mikulla-Roth tiling using molecular dynamics~\cite{koschella_phason_2002} and the analytic polar calculus~\cite{koschella_phason-elastic_2004}. The authors measured the potential energy subjected to phonon strain and -- via approximant boundary conditions -- to phason strain at zero temperature. They showed a quadratic dependence of the energy on strain and determined the generalized elastic constants (phonon, phason, and coupling). One phason elastic constant turned out to be negative, rendering the quasicrystal metastable. This result is in agreement with convex hull calculations of Lee \textit{et al.}~\cite{lee_crystalline_2001} who found only crystalline ground states for binary tilings. The possibility of non-analytic terms has also been documented~\cite{trebin_investigation_2006}, but for these to exist the potential had to show well defined oscillations at more than five next neighbor distances, modeling precisely the matching rules. Such correlated fine tuning of a potential, however, appears most unlikely and renders an energetically stabilized quasicrystal improbable.

As a rule, previous tests of the random tiling hypothesis either (i)~employed discrete lattice models, (ii)~dealt with binary tilings, which are metastable and are complicated by effects of stoichiometry and chemical potential, or (iii)~were restricted to $T=0$. Recently, an interaction pair potential with two minima, the Lennard-Jones-Gauss potential, was proposed~\cite{engel_self-assembly_2007}. At a special choice of positions and depths of the minima, it is possible to grow a stable two-dimensional monatomic decagonal random tiling quasicrystal from the melt in simulations. Thus a simple system close to reality is available, where all types of numerical experiments can be performed to check the hypotheses and paradigms of quasicrystals. Here, we measure the free energy in dependence of temperature and phason strain and derive the temperature dependent phason elastic constants. Our results confirm all the predictions of the random tiling hypothesis: The free energy has gradient-square form, the phason elastic constants increase monotonically with temperature, and close to the point where one of them vanishes, the random tiling turns into a rather complex but periodic approximant crystal. 

\begin{figure}
\centering
\includegraphics[width=0.47\columnwidth]{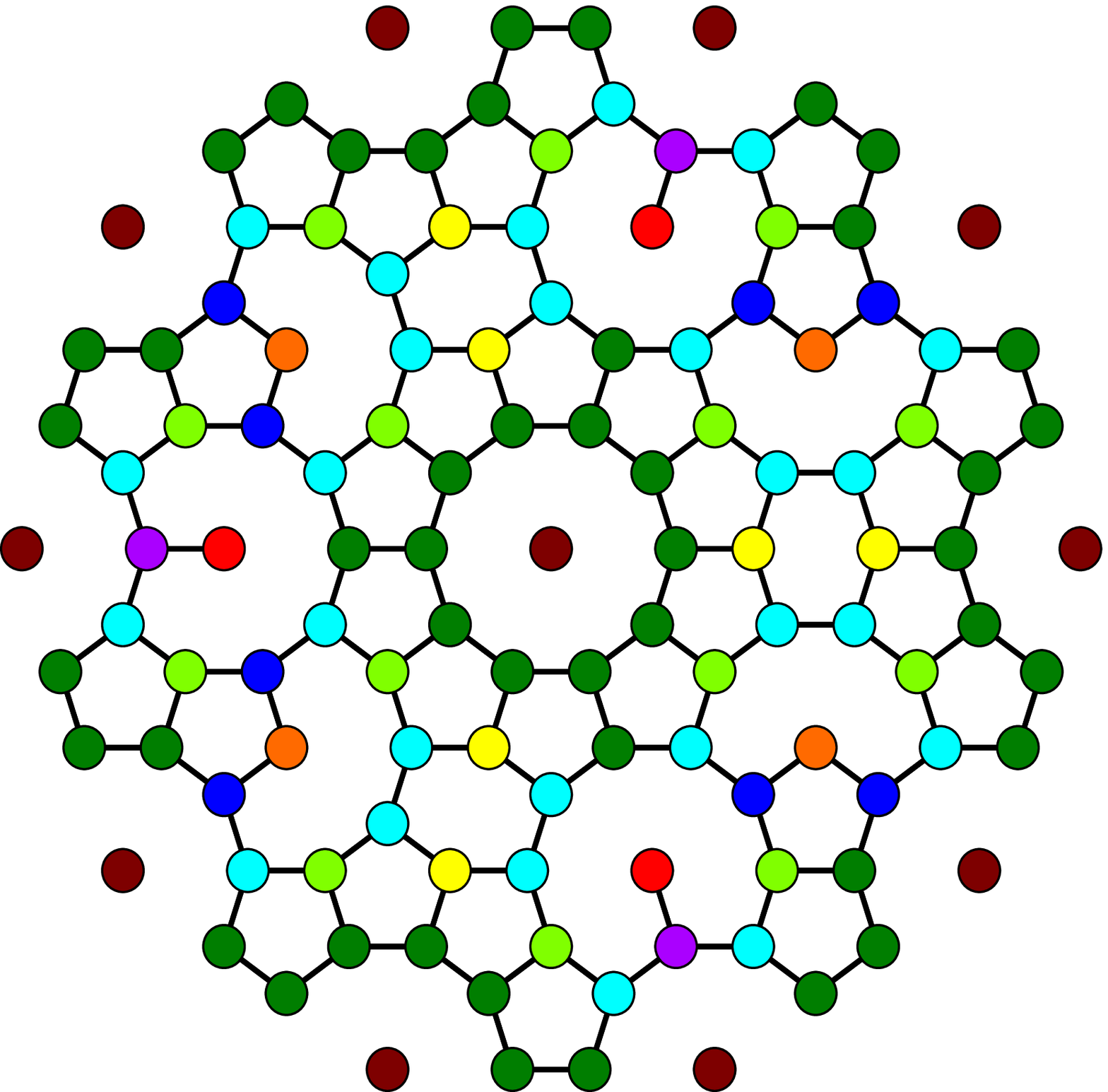}
\hspace{0.02\columnwidth}
\includegraphics[width=0.47\columnwidth]{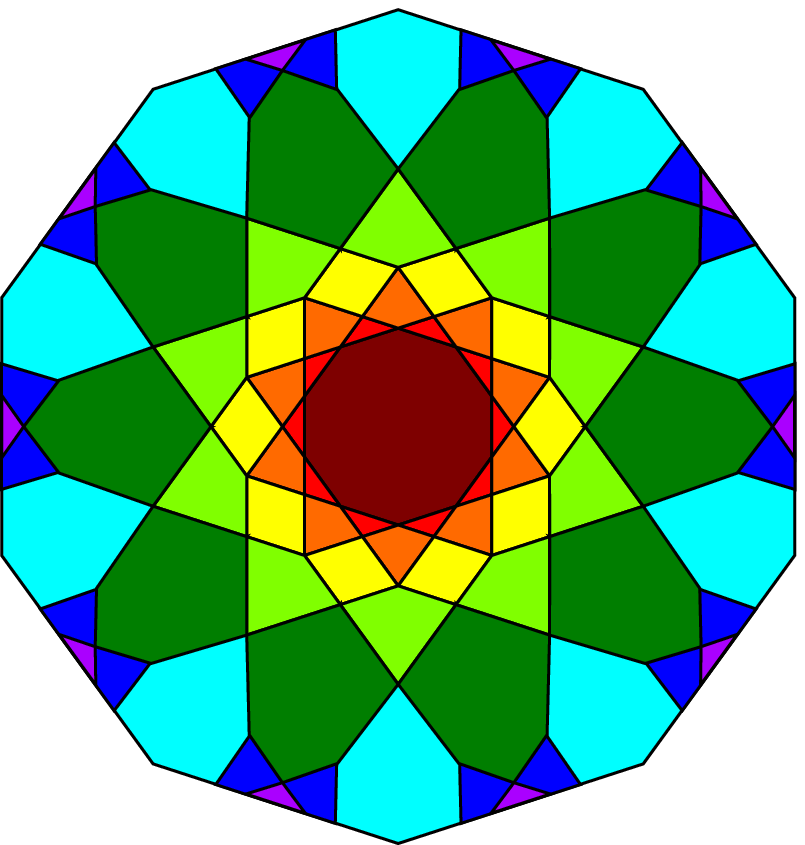}
\caption{Left:~A small patch of the decagonal T\"ubingen triangle tiling. The colors correspond to distinct vertex configurations. Right:~The acceptance domain in perp space is a regular decagon.}
\label{fig:au}
\end{figure}

In the Lennard-Jones--Gauss system identical atoms interact in two dimensions by the simple double-well pair potential
\begin{equation}
V(r) = \frac{1}{r^2} - \frac{2}{r^6} - \epsilon \exp( - \frac{(r-r_0)^2}{2 \sigma^2}).
\end{equation}
A large number of crystal structures is achieved by varying the position $r_0$, depth $\epsilon$ and width $\sigma$ of the second minimum. For the values $r_0= 1.52$, $\epsilon= 1.8$ and $\sigma^2 = 0.02$ and at intermediate temperatures a decagonal random tiling is stabilized~\cite{engel_self-assembly_2007,engel_stability_2008}. It is the random tiling variant of the T\"ubingen triangle tiling~\cite{baake_TTT_1990}, which is composed of thin and thick golden triangles and is related to the Penrose pattern. An alternative representation by polygons (regular and nonconvex decagons, nonagons, hexagons and pentagons) is obtained by connecting nearest neighbors as shown in Fig.~\ref{fig:au}~(left). The T\"ubingen triangle tiling is relevant for modeling experimental quasicrystals including quasiperiodic colloidal monolayers~\cite{Mikhael_colloids_2011}.

\begin{table}
\setlength{\unitlength}{2mm}
\definecolor{AU1}{HTML}{7E0000}
\definecolor{AU2}{HTML}{FF0000}
\definecolor{AU3}{HTML}{FF6A00}
\definecolor{AU4}{HTML}{FFFF00}
\definecolor{AU5}{HTML}{80FF00}
\definecolor{AU6}{HTML}{007E00}
\definecolor{AU7}{HTML}{00FFFF}
\definecolor{AU8}{HTML}{0000FF}
\definecolor{AU9}{HTML}{AA00FF}
\begin{ruledtabular}
\begin{tabular}{ccccccc}
& Vertex  & $\#(r_1)$ & $\#(r_2)$ & $\#(r_3)$ & $E_B$ & $n/N$ \\
\hline
\begin{picture}(2, 1.5)\put(1,0.5){\color{AU1}\circle*{1}}\end{picture} & $V_1$ & 0 & 10 & 0 & -19.6 & $\tau^{-6}$ \\
\begin{picture}(2, 1.5)\put(1,0.5){\color{AU2}\circle*{1}}\end{picture} & $V_2$ & 1 & 9  & 0 & -18.4 & $\tau^{-9}$ \\
\begin{picture}(2, 1.5)\put(1,0.5){\color{AU3}\circle*{1}}\end{picture} & $V_3$ & 2 & 8  & 0 & -17.2 & $2\tau^{-8}$  \\
\begin{picture}(2, 1.5)\put(1,0.5){\color{AU4}\circle*{1}}\end{picture} & $V_4$ & 3 & 7  & 0 & -16.1 & $2\tau^{-7}$  \\
\begin{picture}(2, 1.5)\put(1,0.5){\color{AU5}\circle*{1}}\end{picture} & $V_5$ & 3 & 6  & 1 & -14.5 & $2\tau^{-6}$  \\
\begin{picture}(2, 1.5)\put(1,0.5){\color{AU6}\circle*{1}}\end{picture} & $V_6$ & 3 & 5  & 2 & -12.9 & $4\tau^{-5}$  \\
\begin{picture}(2, 1.5)\put(1,0.5){\color{AU7}\circle*{1}}\end{picture} & $V_7$ & 3 & 4  & 3 & -11.3 & $2\sqrt{5}\tau^{-6}$  \\
\begin{picture}(2, 1.5)\put(1,0.5){\color{AU8}\circle*{1}}\end{picture} & $V_8$ & 3 & 4  & 2 & -10.9 & $4\tau^{-8}$  \\
\begin{picture}(2, 1.5)\put(1,0.5){\color{AU9}\circle*{1}}\end{picture} & $V_9$ & 3 & 3  & 2 & -9.0 & $\tau^{-9}$  \\
\end{tabular}
\end{ruledtabular}
\caption{The vertex configurations $V_i$ of the T\"ubingen triangle tiling, numbers of neighbors in the first three shells of radii $r_1=1$, $r_2=\tau$, $r_3=1.902$, binding energies $E_B$ and frequencies of occurrence $n/N$ in the well-ordered state. Colors are identical to Fig.~\ref{fig:au}. $\tau=(\sqrt{5}+1)/2$ is the golden number.}
\label{tab:vertex}
\end{table}

By applying the cut-off radius $r_c=2.0$, there are nine tiling vertex configurations given in Table~\ref{tab:vertex}. The projections of the vertices into the acceptance domain of perpendicular (perp) space are displayed in Fig.~\ref{fig:au} (right). Note that the potential energy of vertex configurations decreases from the center of the acceptance domain towards the outside with the particles in the center of decagons (dark red) being the most stable. Elementary excitations are local rearrangements of the tiles termed phason flips. The six types of phason flips are listed in Table~\ref{tab:flip}. Only the vertices $V_7$, $V_8$, $V_9$ connected to the boundary of the acceptance domain are able to flip. This is in accordance with the closure condition~\cite{kalugin_closurecondition_1993}, which requires that acceptance domains of random tilings are topologically connected when projected to perp space.

\begin{table}
\begin{ruledtabular}
\begin{tabular}{cccc}
Vertex configuration & Type of flip  & $\varepsilon$  & $E_A$ \\
\hline
$V_9 \rightarrow V_9$ & $F_1$ & 0.0 & 1.89  \\
$V_8 \rightarrow V_9$ & $F_2$ & 1.89 & 3.41 \\
$V_7 \rightarrow V_9$ & $F_3$ & 2.15 & 3.68  \\
$V_8 \rightarrow V_8$ & $F_4$ & 0.0 & 2.89   \\
$V_7 \rightarrow V_8$ & $F_5$ & 0.27 & 3.16   \\
$V_7 \rightarrow V_7$ & $F_6$ & 0.0 & 3.16
\end{tabular}
\end{ruledtabular}
\caption{Phason flips change the vertex configuration of the T\"ubingen triangle tiling. The types of phason flips, difference between final and initial potential energy of the flipping atoms $\varepsilon$, and the height of energy barrier $E_A$ are listed. The energy barriers are calculated here for a single particle motion, but can be lowered significantly in collective motions.}
\label{tab:flip}
\end{table}

We calculate the free energy of the decagonal tiling in dependence of temperature and phason strain $\chi_{ij} = \partial_j w_i$, where $w$ is a two-dimensional phason displacement vector. Following the notation in Refs.~\cite{koschella_phason_2002,koschella_phason-elastic_2004,Henley_randomtilingmodels_1990} and in continuum harmonic approximation, the free phason elastic energy density per particle is 
\begin{align}
F(\chi, T) &= \frac{1}{2} K_2(T) \left((\chi_{6g}^{(1)})^2 + (\chi_{6g}^{(2)})^2 \right) \notag\\
	&+  \frac{1}{2} K_1(T)  \left((\chi_{8g}^{(1)})^2 + (\chi_{8g}^{(2)})^2 \right),
\label{eq:phel}
\end{align}
with symmetrized phason strain modes 
\begin{equation}
\left(\begin{array}{c} 
\chi_{6g}^{(1)} \\ 
\chi_{6g}^{(2)} \\ 
\chi_{8g}^{(1)} \\ 
\chi_{8g}^{(2)}
\end{array} \right) = \left(
\begin{array}{cccc}
1 & -1 & 0 & 0 \\ 
0 & 0 & -1 & -1 \\ 
1 & 1 & 0 & 0 \\ 
0 & 0 & 1 & -1
\end{array} \right)
\left(
\begin{array}{c} 
\chi_{11} \\ 
\chi_{22} \\ 
\chi_{12} \\ 
\chi_{21}
\end{array} \right).
\label{eq:strain}
\end{equation}

The phason elastic constants $K_2$ and $K_1$ are temperature dependent and are weighting strain contributions of the $\Gamma_{6g}$ and $\Gamma_{8g}$ irreducible representations of the symmetry group $D_{10h}$. Both types of strain allow a symmetry breaking to periodic approximants of orthorhombic $C_{2v}$ symmetry, as they, when subduced, contain the trivial $A_1$-representation ($\Gamma_{6g}|C_{2v} = A_1 \oplus A_2 = \Gamma_{8g}|C_{2v}$). In our group theory notation we follow Ref. \cite{authier_international_tables_2003}.

The presence of phason strain is necessary for the transformation of a quasicrystal into a periodic approximant. In turn, by constructing a specific approximant in simulation using periodic boundary conditions, a well-defined average phason strain can be enforced artificially. We have constructed a series of orthorhombic approximants with in the order of 20000 particles and phason strain $|\chi_{11}|, |\chi_{22}|\leq 0.2, \chi_{12} = \chi_{21} = 0 $ following~\cite{koschella_phason_2002}. Thus, only $\chi_{6g}^{(1)}$ and $\chi_{8g}^{(1)}$ were non-zero, and by varying them we can determine $K_1$ and $K_2$.

It has been shown that typical phonon frequencies are at least two orders of magnitude larger than phason flip rates~\cite{engel_dynamics_2010}. Since the time scales of the two excitations are well decoupled, we can separate the free energy via $F=F_\text{phon}+F_{\text{conf}}$ into a phonon part accessible to evaluation by molecular dynamics and a configurational part due to the tiling degeneracy caused by phason flips, where we must apply accelerated methods.

We calculate the phonon part with a combination of the Frenkel-Ladd method~\cite{frenkel_new_1984} and thermodynamic integration as in~\cite{engel_entropic_2011}. At the low temperature $T_0=0.05$, the absolute free energy is determined with Frenkel-Ladd by interpolating between the target structure and an Einstein crystal with known free energy. It is then extended to higher temperature by integrating over the internal energy,
where we take the dependence on phason strain explicitly into account:
\begin{equation}
F_{\text{phon}}(\chi,T) = E_0(\chi) +T g(\chi) + I(\chi,T).
\label{eq:freeenergy}
\end{equation}
Here $g(\chi)$ is an integration constant obtained by comparison with Frenkel-Ladd and $E(\chi,T)$ the internal energy per particle with ground state potential energy $E_0(\chi)=E(\chi,0)$. Thermodynamic integration gives
\begin{equation}
I(\chi,T) = - T \int_{T_0}^T \mathrm{d}T'\, \frac{E(\chi, T')-E_0(\chi)}{T'^2}.
\label{eq:I}
\end{equation}

For the configurational part, we apply an approximation of uncorrelated flips. A justification for this approximation will be given below. At each value of the phason strain we determine the vertices with vertex configuration $V_7$, $V_8$, or $V_9$. Only these vertices are able to flip without introducing tiling defects and can be associated with a flip type $F_k$ (see Table \ref{tab:flip}). A simple Ansatz is an Ising-type model with Hamiltonian
 \begin{equation}
H(\{s_{ki}\}) = \sum_{k=1}^6 \sum_{i=1}^{n_k}\varepsilon_k s_{ki}
\end{equation}
using the flip energies $\varepsilon_k$ (Table~\ref{tab:flip}) between initial state $s_{ki}=0$ and final state $s_{ki}=1$, and the number of flips per flip type $n_k$. The configurational free energy calculated from the canonical partition function is
\begin{equation}
F_{\text{conf}}(\chi,T)= -k_BT \sum_k  \frac{n_k(\chi)}{N(\chi)} \ln(1 + e^{-\varepsilon_k/k_BT}).
\label{eq:config}
\end{equation}
Here $N(\chi)$ is the number of particles in the unit cell of the approximant. Note that only the numbers $n_k$ depend on phason strain, but not the flip energies. In order to determine the free energy as a function of phason strain we measure $n_k$ for  each approximant. Finally, combining the phonon and configuration part of the free energies the phason elastic constants are given by the curvature at zero phason strain
\begin{equation}
K(T) = \frac{\partial^2 F(\chi,T)}{\partial \chi^2}
\label{eq:phasel}
\end{equation}
where $\chi = \chi_{6g}^{(1)}$ for $K_2$ and $\chi = \chi_{8g}^{(1)}$ for $K_1$. 

\begin{figure}
\centering
\includegraphics[width=\columnwidth]{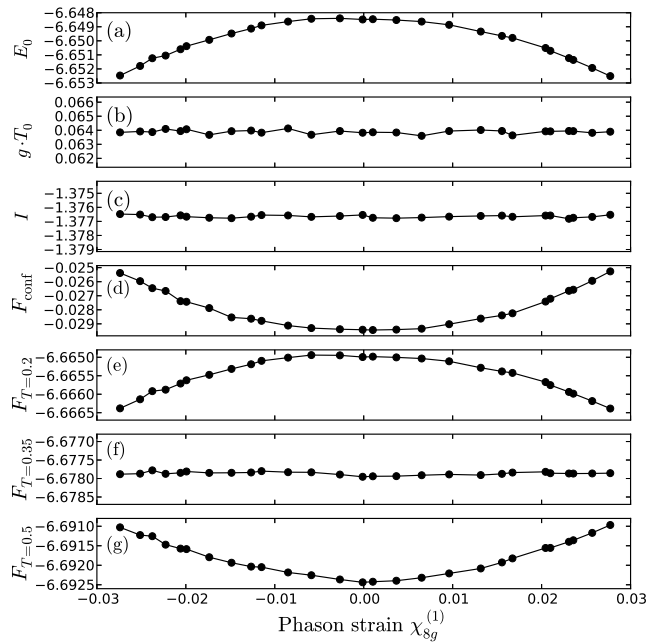}
\caption{Contributions to the free energy for phason strain of symmetry $\Gamma_{8g}$. (a) Ground state potential energy $E_0(\chi)$, (b) the Frenkel-Ladd integration constant $g(\chi)\cdot T_0$, (c) the thermodynamic integral $I(\chi,T)$ and (d) the configurational part $F_{\text{conf}}(\chi,T)$ at temperature $T=0.35$. (b) and (c) are phason strain independent and do not contribute to the phason elastic constant. (e)-(g) The free energy at three different temperatures.}
\label{fig:fe}
\end{figure}

Fig.~\ref{fig:fe} (a)-(d) presents the different contributions to the free energy at $T = 0.35$ for phason strain of symmetry $\Gamma_{8g}$. Phason strain of symmetry $\Gamma_{6g}$ behaves similarly (not shown). The part $Tg(\chi)$, which arises from the numerically costly Frenkel-Ladd method, is large and shows strong fluctuations, but is practically strain independent. In addition, the part $I(\chi,T)$, which contains the vibrational entropy, is also not influenced by phason strain. This means only the terms $E_0(\chi)$ and $F_{\text{conf}}(\chi,T)$ contribute to Eq.~(\ref{eq:phasel}). They are parabolic and we have $K_i(T) = K_{i,\text{phon}} + K_{i,\text{conf}}(T)$. The temperature dependence is contained completely in the configurational part.

\begin{figure}
\centering
\includegraphics[width=\columnwidth]{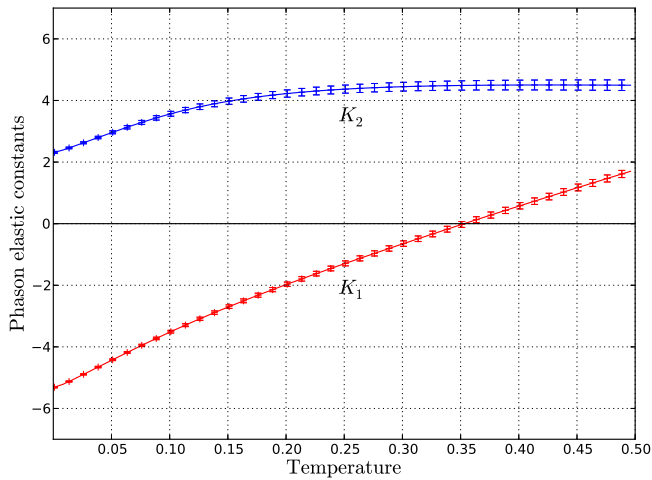}
\caption{Temperature dependence of the phason elastic constants $K_1$ and $K_2$. At low temperature, $K_1$ is negative and $K_2$ is positive. Both phason elastic constants grow linearly at low temperature. $K_1$ changes sign at $T_{1c}=0.35 \pm 0.01$. Error bars are calculated as the regression error for the parabolic approximation of the free energy Eq.~(\ref{eq:phel}).}
\label{fig:phasel}
\end{figure}

At low temperature, the configurational free energy is dominated by the degeneracy of the ground state (caused by the flips $F_1$, $F_4$, $F_6$, see Table \ref{tab:flip}). The phason elastic constants are linear in temperature, $K_{i,\text{conf}}(T) \propto T$ \cite{widom_entropic-qc_1989}. We observe that the ground state potential energy $E_0(\chi)$ has a saddle point at $\chi=0$ and the phason elastic constants are $K_1<0$ and $K_2>0$ (Fig.~\ref{fig:phasel}), which demonstrates that the random tiling decagonal quasicrystal is not stable at $T=0$. With rising temperature the configurational entropy increases both elastic constants. $K_1$ changes sign at $T_{1c}=0.35 \pm 0.01$. The reversal of the curvature of the total free energy is shown for three temperatures in Fig.~\ref{fig:fe} (e)-(g). In agreement with this results and close to this point, at $T_{2c}=0.37 \pm 0.03$, MC simulations of Ref.~\cite{engel_self-assembly_2007} had already shown a phase transition from a periodic Xi-phase to the quasicrystal. The Xi-phase is a (3/2, 2/1) approximant, stabilized by the maximization of energetically favored centered decagons, the lowest energy vertex ($V_1$) in Table~\ref{tab:vertex}.

\begin{figure}
\centering
\includegraphics[width=\columnwidth]{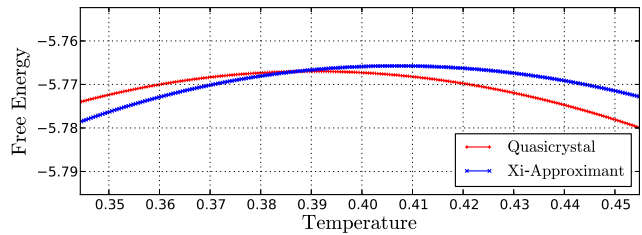}
\caption{The free energies of the quasicrystal and the Xi-approximant cross at $T_{3c}=0.39$.}
\label{fig:phasetr}
\end{figure}

As a separate test, we calculated the free energy difference between the Xi-phase and the unstrained decagonal quasicrystal directly. The internal energy difference $E(\chi, T)-E_0(\chi)$ was measured with molecular dynamics to high precision and fitted using a power series up to third order in $T$. This allowed solving the integral in Eq.~(\ref{eq:I}) analytically. We find that the free energies of both phases cross at $T_{3c}=0.39$ (Fig.\ref{fig:phasetr}), which confirms once more the presence of a crystal-quasicrystal transition.

The behavior we observe resembles X-ray investigations of icosahedral Al-Cu-Fe by Bancel \textit{et al.}~\cite{bancel_dynamical_1989}, where decreasing diffraction peak intensities and a structural transition to a periodic phase are explained by the softening of a phason elastic constant. In principle, a softening transformation can induce a single transition directly to a low temperature approximant or a chain of transitions by running through all orders of rational approximants~\footnote{A continuous transition from an aperiodic phase to a periodic phase is known as the devil's staircase in incommensurately modulated structures, see~\cite{Steurer_phasetransformations_2005, engel_entropic_2011}.}. So far, the latter has not been observed in experiment. The crossing of the free energies of Xi-phase and decagonal quasicrystal in Fig.\ref{fig:phasetr} points to a first order transition. A possible chain of approximants in between might have been suppressed because we were forced to use periodic boundary conditions. In the MC simulations of Ref.~\cite{engel_self-assembly_2007}, open boundaries were used. There, indeed, the phason strain increases monotonically from zero to the value of the Xi-approximant with falling temperature, but the investigated systems were too small to allow tracking of large approximants.

It remains to justify the approximation of uncorrelated phason flips. We observe in the completely relaxed quasicrystal that only flips of types $F_1$ are possible. They do not cost any energy, are isolated and uncorrelated. Correlated flips (and eventually diffusion) are only possible in combination with flips of other types, which partly have a finite energy penalty (see Table \ref{tab:flip}) and are much less probable~\cite{engel_dynamics_2010}. Hence, the approximation of uncorrelated flips is exact for the quasiperiodic ground state, and the low-temperature behavior of the phason elastic constants is correct to first order. In his seminal work on random tilings Henley commented on the existence of "a very close correlation between the number of flippable sites and the entropy per node"~\cite{Henley_randomtilingmodels_1990}, a correlation that is analogue to our approximation of uncorrelated flips. The agreement of $T_{3c}=0.39$, obtained by free energy calculations, and the MC result $T_{2c}=0.37 \pm 0.03$ strengthens our confidence in the approximation of uncorrelated phason flips further. We note that the approximation does not affect the confirmation of the random tiling hypothesis nor any of our conclusions.

The two-dimensional fully atomistic system studied in this work represents a most simple nontrivial quasicrystal. It allows to extract features which are due solely to quasiperiodicity. Two terms compete: (i)~potential energy and phonon motion create a saddle point in the phason strain enforcing periodicity at low temperature; (ii)~configurational entropy favors quasiperiodicity and dominates at high temperature. Our results agree with the predictions for a maximally random tiling~\cite{Henley_randomtilingmodels_1990}, although the assumption of energetic degeneracy of all tiling configurations is not necessary.

\bibliography{kiselev-12-bibtex}

\end{document}